\documentclass[journal=jacsat,manuscript=article]{achemso}
\usepackage[version=3]{mhchem} 
\usepackage[utf8]{inputenc}
\usepackage[T1]{fontenc}
\usepackage{wrapfig}
\usepackage{amsmath}
\usepackage{multirow}
\usepackage{multicol,graphicx,float}
\usepackage{xcolor}
\usepackage{ulem}

\title{Synthesis of hybrid gold-protein nanobioconjugates with strongly enhanced circular dichroism}

\author{Yoann Prado}
\affiliation{Sorbonne Universit\'{e}, CNRS, Institut des NanoSciences de Paris, INSP,  F-75252 Paris, France}
\author{Victoire Ho}
\affiliation{Sorbonne Universit\'{e}, CNRS, Institut des NanoSciences de Paris, INSP,  F-75252 Paris, France}
\author{Maud Virolle}
\affiliation{Sorbonne Universit\'{e}, CNRS, Institut des NanoSciences de Paris, INSP,  F-75252 Paris, France}
\author{Maria Sanz-Paz}
\affiliation{Sorbonne Universit\'{e}, CNRS, Institut des NanoSciences de Paris, INSP,  F-75252 Paris, France}
\author{Mathieu Mivelle}
\affiliation{Sorbonne Universit\'{e}, CNRS, Institut des NanoSciences de Paris, INSP,  F-75252 Paris, France}
\author{Souhir Boujday}
\affiliation{Sorbonne Universit\'{e}, CNRS, Laboratoire de R\'eactivit\'e de Surface, LRS,  F-75252 Paris, France}
\author{Michèle Salmain}
\affiliation{Sorbonne Universit\'{e}, CNRS, Institut Parisien de Chimie Mol\'eculaire, IPCM,  F-75252 Paris, France}
\author{Guillaume Demesy}
\affiliation{Aix-Marseille Universit\'{e}, CNRS, Centrale Marseille, Institut Fresnel, 13013 Marseille, France}
\author{Nicolas Bonod}
\affiliation{Aix-Marseille Universit\'{e}, CNRS, Centrale Marseille, Institut Fresnel, 13013 Marseille, France}
\author{Bruno Gallas}
\affiliation{Sorbonne Universit\'{e}, CNRS, Institut des NanoSciences de Paris, INSP,  F-75252 Paris, France}
\email{bruno.gallas@sorbonne-universite.fr}

\begin{document}

\begin{abstract}
\textit{Controlling chiral light-matter interactions with nanophotonics is a powerful strategy for amplifying molecular circular dichroism (CD) beyond its natural limits. Here, we present a plasmonic platform that uses superchiral near fields to amplify the optical activity of biological chromophores. Gold nanoparticles ($\approx$ 3.6 nm in diameter) are electrostatically coupled with photosynthetic proteins which results in stable hybrid nanobioconjugates in aqueous solution. The Q-band absorption spectrum of the proteins overlaps with the localized surface plasmon resonance of the gold particles. In the visible region, a combination of numerical simulations and normalized CD measurements reveal an enhancement factor of 3 compared to free proteins. This observation reveals three effects. First, there is a local increase in electromagnetic chirality density in the protein environment. Second, there is an increase in the absorption of the proteins. Third, there is plasmon-induced circular dichroism.  Our results quantitatively demonstrate that near-field super-chirality directly modulates biomolecular optical activity. These findings open new avenues for chiroptical nanodevices, biosensing platforms, and light-driven asymmetric photochemistry.}
\end{abstract}

\section{Keywords} 
Plasmonics; Nanobioconjugates; Circular Dichroism; Phycobiliproteins; Numerical Simulations 

\section{INTRODUCTION}
All biological macromolecules are chiral, meaning that they cannot be superimposed to their mirror image. Their biochemical functionality is intimately linked to this handedness, which depends on the absolute configuration of stereocenters or the secondary structure of the proteins. Conventional chiroptical techniques rely on the differential absorption of right- and left-circularly polarized light, known as circular dichroism (CD). However, CD signals are typically only on the order of $10^{-4}$ of the total absorption, which severely limits measurement sensitivity. Enhancing light–matter interactions using near-field concentration provided by photonic resonators, particularly plasmonic resonators, has transformed numerous biosensing approaches. Yet, achieving a comparable enhancement for chiral detection remains challenging. Following the pioneering work of Tang and Cohen, it is now admitted that the relevant quantity governing chiral light–matter interactions is the optical chirality density $C=-\frac{\epsilon_o \omega}{2}Im(\textbf{E$^*$}.\textbf{B})$.\cite{TangCohen2010, Alejandro_2018}

This framework has motivated the design of chiral plasmonic architectures engineered to boost the local chirality density, giving rise to so-called superchiral fields.\cite{Hendry2013, Kumar2019, Alu2017, Hendry2010, PaivaMarques2020_ChiralPlasmonics_Sensors, Weiss_ACS_2016, Link2019, Genet2014, Kadodwala_2020} Metasurfaces composed of resonators with opposite handedness have enabled differential detection schemes, though their performance is often hindered by the highly nonuniform spatial distribution of the chiral near fields.\cite{Naik2010, Weiss2022, Adhikari2022} Achiral resonators have also been shown to modify the CD response of adjacent molecules.\cite{Oleg_NanoLett_2013, Nicolas2025, Davis2012, Hendry2013, Giessen2012, Cui2023, Nicolas2023} Early studies on gold disks revealed CD signatures at the plasmon resonance,\cite{Naik2010,DiskGovorov_2013} initially attributed to dipole–dipole coupling between the plasmonic electric dipole and the molecular electric and magnetic dipoles.\cite{Naik2010, Naik2011, DiskGovorov_2013, Nesterov2016, García-Guirado2020} Subsequent work has identified additional contributions, including multipolar interactions, near-field phase structure, and hybrid plasmon–exciton effects.\cite{Multipole_PRB2008, Efrima_1983, Efrima_1984, Rho2020, Weiss2022, Nesterov2016, MunRho_2019,Fang_2025,Link2019} CD enhancement has also been demonstrated, both numerically and experimentally, for dielectric resonators coated with amino-acid thin films.\cite{García-Guirado2020}

Surface-based measurements, however, introduce further complexity due to artifacts that obscure the intrinsic molecular response.\cite{Arteaga2016, McPeak2019, Nicolas2023, OMarkovic2025} Colloidal nanoparticle suspensions have thus emerged as attractive sensing media, and transfer of optical chirality from chiral analytes to the Localized Surface Plasmon Resonance (LSPR) or to excitonic transitions has been reported.\cite{Naik2011, C1JM12345A, Gunko2007} Nevertheless, CD features near the LSPR have frequently been attributed to nanoparticle morphology or to emergent chirality in nanoparticle assemblies rather than to genuine molecular signals.\cite{Parola_2022, Marzan_2020, Link2019, Schreiber2013_DNA_Nanostructures_CD} These ambiguities arise in part because many biomolecules lack CD-active transitions at the plasmon resonance frequency, leading to interactions mediated only by the weak chiral absorption tails of the analyte. DNA-based dyes have been proposed to probe the effect of resonant coupling between gold nanoparticles and chiral absorption.\cite{LinkDNA2019} A complex change in the CD lineshape was observed and was tentatively attributed to a combination of plasmon-induced CD and mode hybridization.\cite{Naik2010,LinkDNA2019}

Here, we address this issue by investigating biomolecular systems that do possess strong CD-active absorption bands resonant with plasmonic nanoparticles dispersed in solution. We focus on phycobiliproteins (PBPs), that are highly efficient light-harvesting complexes found in cyanobacteria and red algae. R-phycoerythrin (R-PE), in particular, is a major photosynthetic pigment composed of two different polypeptide chains $\alpha$ and $\beta$ regrouped in a structure that incorporates phycoerythrobilin and phycourobilin chromophores.\cite{DAGNINOLEONE20221506} These pigments absorb in the blue–green to yellow–green region and fluoresce near 575–580\,nm with quantum yields of $\sim$0.84–0.85.\cite{NewRPE_1996,Bunster2004} Owing to their strong optical response and facile bioconjugation, PBPs are widely used in immunofluorescence, flow cytometry, and bioimaging.\cite{DAGNINOLEONE20221506,Review_algae,JPhysChem_Andrew_2011} Importantly, the excitonic Q-band of R-PE spectrally overlaps with the LSPR of colloidal gold nanoparticles in water, enabling true resonant coupling which could be extended to asymmetric chemistry or catalysis.

We show that R-PE molecules positioned within the near field of gold nanoparticles exhibit a pronounced, 3-fold enhancement of their CD signal. Numerical simulations reveal that this enhancement originates from two synergistic effects: a substantial increase in the local optical chirality density and in the absorption of biomolecule, and a transfer of optical chirality between the biomolecule and the nanoparticle. Together, these mechanisms provide a unified description of chirality amplification in resonantly coupled biomolecule–nanoparticle systems. Numerical simulations show that our results may be extended to any free-electron metal even outside the LSPR, although LSPR may provide 10-fold enhancement.

\section{RESULTS}

\subsection{Synthesis of gold nanoparticles and AuNP/R-PE nanobioconjugates}

\begin{figure*}[!ht]
\centering\includegraphics[width=0.7\linewidth]{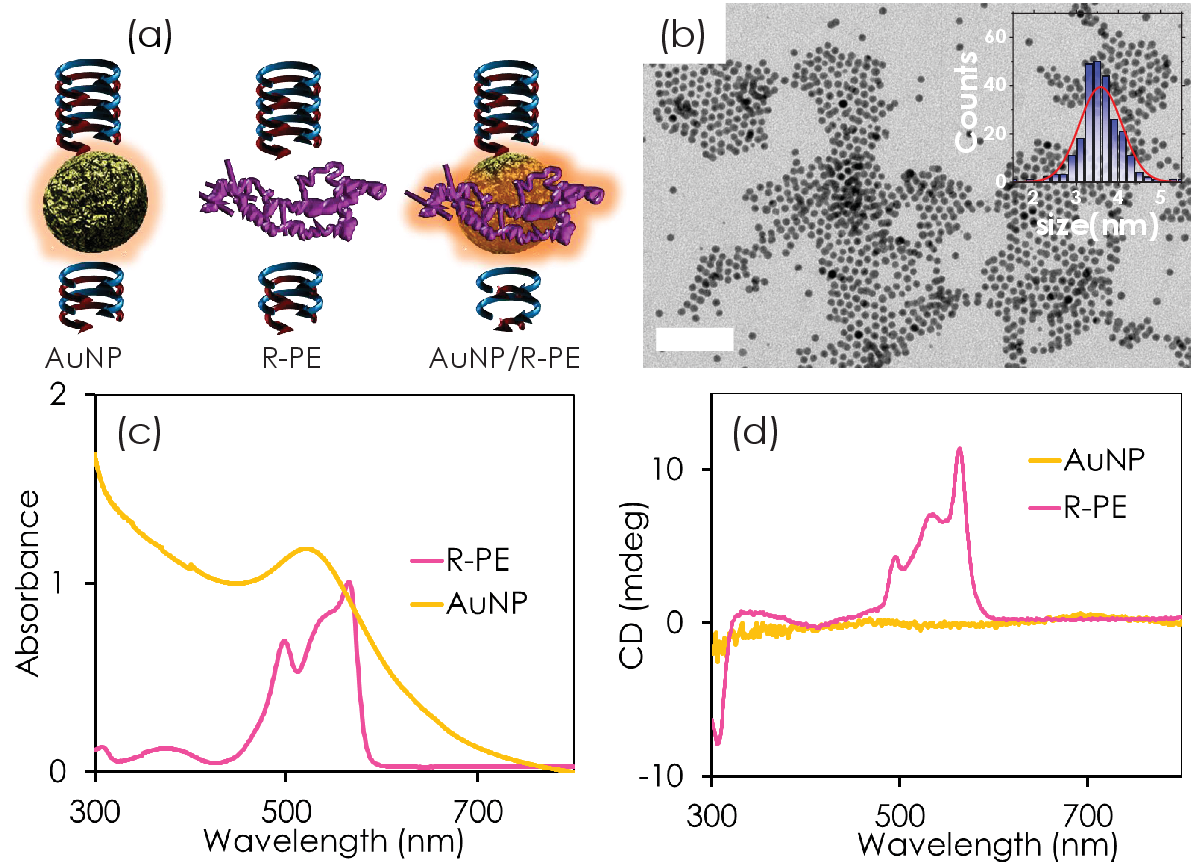}
\caption{(a) Schematic illustration of protein CD enhancement arising from the increased optical chirality density in the near field of a gold nanoparticle. (b) Transmission electron micrograph of the synthesized AuNP. Insert: size distribution and gaussian fit (<D>=3.6nm, $\sigma$=0.5nm). (c) Measured absorption spectra of the AuNP colloidal suspension in water and of R-PE in PBS, highlighting the spectral overlap between the plasmon resonance and the R-PE Q-band. (d) Circular dichroism spectra measured for the same AuNP and R-PE solutions.}
\label{fig:concepts}
\end{figure*}

 Figure \ref{fig:concepts}(a) illustrates the concept of protein CD enhancement mediated by gold nanoparticles (AuNP). Under circularly polarized illumination, the near field generated at the plasmon resonance of an AuNP exhibits an increased optical chirality density $C$ and absorption, even though the nanoparticle itself does not produce a detectable far-field CD signal. A protein in solution, when excited by circularly polarized light, absorbs differentially owing to the chiral density of the incident field, which is bounded by unity for a plane wave, resulting in a weak CD response. When the same protein is positioned within the plasmonic near field, it experiences a local optical chirality significantly larger than that of the incident wave, leading to a correspondingly enhanced far-field CD.
 
 The colloidal AuNP used in this work were synthesized to yield particles with an average diameter of 3.6\,nm and a size dispersion of $\pm 0.5$\,nm, as shown in the transmission electron micrograph in Figure \ref{fig:concepts}(b). Their optical extinction spectrum, displayed in Figure \ref{fig:concepts}(c), exhibits a broad plasmon resonance centered at 511\,nm. In other hand, the phycobilin chromophores in R-PE, conserved in phosphate buffered saline (PBS, pH=7.2) show electronic transition in the visible, in the 500-600 nm range. This excitonic absorption spectrally overlaps with this plasmon band.

Figure \ref{fig:concepts}(d) presents the CD spectra of the same AuNP suspension in water (pH=7.3) and of R-PE in PBS. As expected, no CD signal is observed from the AuNP. In contrast, the CD of R-PE shows its characteristic three positive bands in the 500–600\,nm range and a negative band near 300\,nm. The band at 570\,nm reaches 11.4\,mdeg for a solution with an absorbance of 1 at this wavelength, consistent with the CD lineshape of R-PE.\cite{NewRPE_1996, Kovalev2013} The corresponding $g$-factor at 570\,nm, determined for R-PE concentrations between 0.161\,$\mu$M and 5\,$\mu$M, averaged to $3.5\times 10^{-4}$, in excellent agreement with literature values.\cite{NewRPE_1996} 

\begin{figure*}[!ht]
\centering\includegraphics[width=0.9\linewidth]{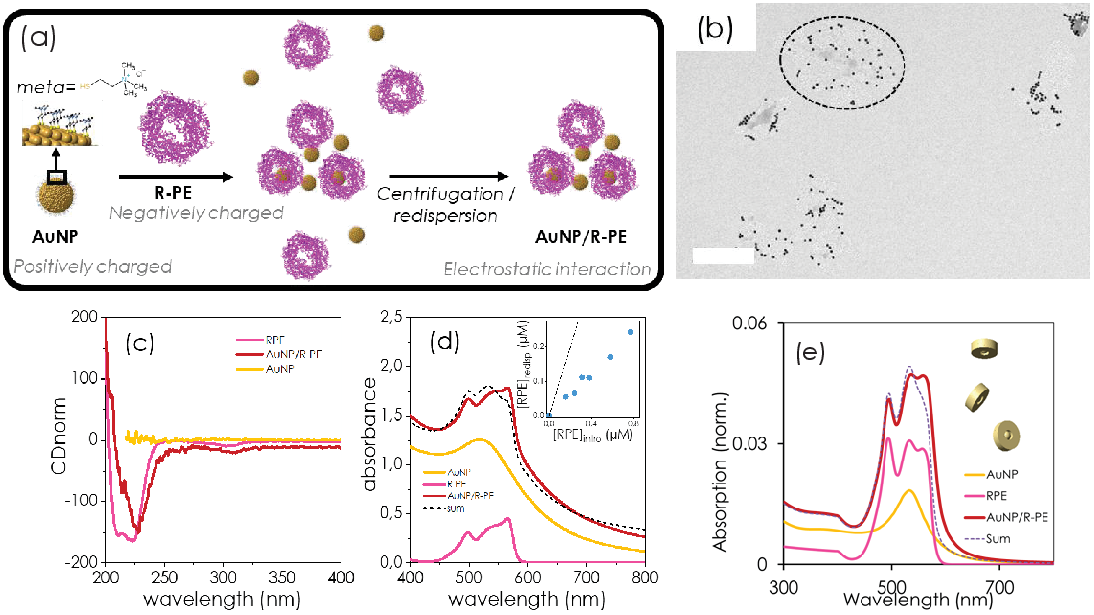}
\caption{(a) Schematics of the nanobioconjugate and purification process. (b) TEM micrograph of AuNP/R-PE nanobioconjugates (scale bar 200nm). The dotted line delineates one nanobioconjugate (c) CD in the UV range of R-PE (pink), AuNP (yellow) and purified AuNP/R-PE (red) showing the evolution of the secondary structure of R-PE in the nanobioconjugates. (d) Measured and (e) calculated absorption spectrum of AuNP (yellow), R-PE (pink) and purified AuNP/R-PE (red) and sum of the separate contributions AuNP and R-PE (black dotted line). Inset: R-PE concentration in the redispersion [R-PE]$_{redisp}$ as a function of the R-PE concentration introduced initially on the AuNP dispersion [R-PE]$_{intro}$  (dark line stands for the full integration of the proteins on the AuNP).}
\label{fig:assemblies}
\end{figure*}

The coupling between AuNP and R-PE was obtained through electrostatic assembling yielding nanobioconjugates (Figure \ref{fig:assemblies}(a)). AuNP prepared from Ref. \cite{AuNP_Peng} were initially dispersed in toluene with oleylamine ligands at the nanoparticle surface. AuNP were then transferred to water thanks to the exchange of oleylamine ligand by the positively charged 2-mercaptoethyl-N,N,N-trimethylammonium, leading to  positively charged AuNP.  Since its isoelectric point is 4.8, R-PE is negatively charged in PBS at pH=7.2. The positive and negative charge of AuNP and R-PE respectively has been confirmed by zeta potential measurements (see Figure S1 and Table 1 in S.I.). The nanobioconjugates were obtained by adding different volumes of R-PE, leading to different ratio [R-PE]/[AuNP] varying from 0.05 to 0.14 (see Table 2 in S.I.). The influence of the electrostatic interaction on the assembling mechanism has been clearly identified by monitoring the influence of the AuNP's charge on the formation of nanobioconjugates after mixing the AuNP and the R-PE solutions (see Figure S2). The solution was then purified by a centrifugation step, in order to precipitate the AuNP/R-PE nanobioconjugates only. The supernatant containing residual R-PE and AuNP was removed and the precipitate was redispersed in 3mL of PBS (see Figure S3 in S.I.). 
TEM image of a dropcast of the dispersion on a TEM grid is shown in Figure \ref{fig:assemblies}(b). Because of the low contrast of proteins in electron microscopy, the R-PE is observed only as a darker area containing the well-defined AuNP. This image was very different from the one obtained for AuNP only as presented in Figure \ref{fig:concepts}(b). Large aggregates were observed which may occur from a combination of charge compensation in nanobioconjugates (see Figure S1), of aggregation during the purification step and of TEM preparation. 

The proteins observed by TEM exhibited insufficient contrast to confirm the structural integrity of R-PE. The secondary structure of proteins may be significantly altered when confined near nanoparticle surfaces. To assess potential structural modifications, CD spectra in the ultraviolet range (200--400\,nm) were recorded for the R-PE solution and for the purified nanobioconjugates, as shown in Figure~\ref{fig:assemblies}(c). As expected, AuNP exhibited no spectral features in this region. In contrast, R-PE in PBS displayed mostly characteristic signatures of $\alpha$-helices with minima at 210 nm and 218 nm as already reported.\cite{LIU2009, Kovalev2013,CHANG1996} The CD spectrum of the purified nanobioconjugates with a ratio [R-PE]/[AuNP]=0.14 is also presented. The secondary structure of the R-PE in AuNP/R-PE was different from that of R-PE in PBS (see Figure~\ref{fig:assemblies}(c) and Figure S4 in S.I. for other ratio) with a significant reduction of the contribution near 210 nm. Modifications of the relative intensity of this signatures have been observed where R-PE was used as reactor for the synthesis of CdS or Ag$_2$S quantum dots suggesting that the $\alpha$-helices partly uncoil upon interaction with nanoparticles.\cite{Kovalev2013, Review_tunnel_cavity} 

The purified nanobioconjugates were characterized using UV-vis absorption. A typical absorption spectrum with a ratio [R-PE]/[AuNP]=0.14 is presented in Figure \ref{fig:assemblies}(d) (see also Figure S5 for other ratio). It can be seen that the signatures of the plasmon resonance of the AuNPs and of the Q-band transitions of the R-PE are superimposed. The absorption spectra of all as-prepared nanobioconjugates, supernatants and redispersions are presented in Figure S3. The measured absorption spectra were decomposed into contributions from AuNPs and R-PE based on their respective reference spectra (Figure \ref{fig:concepts}(c)), as illustrated in Figure~\ref{fig:assemblies}(d) and Figure S5 and Table 2 in SI. The relative contributions were used to estimate the composition of each purified nanobioconjugate as a function of the initial R-PE concentration added to the AuNP suspension. The inset of Figure~\ref{fig:assemblies}(d) shows the variation of the R-PE concentration in the redispersion which increased with the R-PE concentration introduced initially on the AuNP dispersion. No saturation was observed in agreement with the low ratio of the R-PE/AuNP (<0.14). It should be noted that the simple linear superposition of AuNP and R-PE spectra did not perfectly reproduce the measured absorption, particularly near 570\,nm. This deviation could be expected, as the local dielectric environment of AuNPs within the nanobioconjugates differs significantly from that of pure water. Numerical simulations were performed to confirm that strong spectral reshaping occurred near 570\,nm, they are presented in Section 'Numerical Analysis' hereafter. They show a similar optical response associated with the plasmon resonance as shown in Figure~\ref{fig:assemblies}(e) (see also Figure S6). An enhancement of the absorption of the proteins reaching a peak value of 1.45 near 588 nm was observed (Figure S6). However, the close agreement between the simulated and experimentally decomposed spectra supports the validity of the decomposition method employed here, keeping in mind that the actual R-PE concentration may be slightly over-estimated.

\subsection{Analysis of the CD enhancement in the nanobioconjugates}
\begin{figure*}[!hb]
\centering\includegraphics[width=0.8\linewidth]{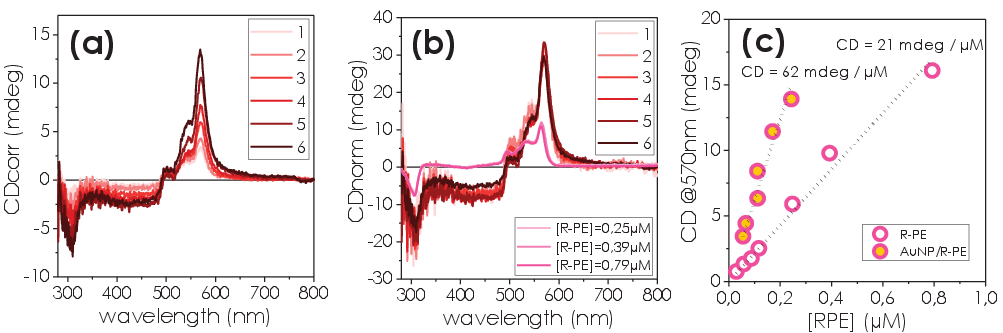}
\caption{CD enhancement of R-PE in AuNP/R-PE nanobioconjugates. (a) Raw CD spectra in the Trp-band and Q-band regions for purified nanobioconjugates prepared with different initial R-PE concentrations. (b) CD spectra for all nanobioconjugates normalized such that the R-PE absorption at maximum, near 570\,nm, is equal to unity. The CD spectra of R-PE in PBS similarly normalized are plotted in blue. (c) CD amplitude at 570\,nm as a function of the initial R-PE concentration. A linear dependence is observed in both cases, with slopes of 21\,mdeg$/\mu$M for R-PE in PBS and 62\,mdeg$/\mu$M for the AuNP/R-PE nanobioconjugates.
}
\label{fig:CDRPE}
\end{figure*}

The influence of the enhanced chiral near field on the CD response of R-PE within the AuNP/R-PE nanobioconjugates is expected to be most pronounced near the plasmon resonance of the AuNPs. Figure~\ref{fig:CDRPE}(a) presents the CD spectra recorded from 250 to 800 nm for the purified nanobioconjugates as a function of the initial R-PE concentration. As expected, the overall CD amplitude increased with increasing R-PE content. Despite this amplitude change, the spectral lineshape remained characteristic of R-PE in PBS, with a negative band near 300 nm associated with the contribution of tryptophans in R-PE (Trp-band) and three positive bands (Q-band) between 500 and 600\,nm. A weak negative plateau was observed in the 350-500\,nm region; this feature may partly arise from subtraction of the background CD signal of the AuNP suspension (see Figure S7). However, as shown in the next section, it may also reflect a genuine signature of the R-PE–AuNP coupling. The position of the maximum CD near 570 nm was also slightly red-shifted (see Figures S3 and S8) for all nanobioconjugates.

To enable a direct comparison between the CD of R-PE in PBS and that of R-PE within the purified AuNP/R-PE nanobioconjugates, all CD spectra were normalized such that the R-PE absorption at maximum near 570\,nm, as determined in Figure \ref{fig:assemblies}(d), was set to unity. The resulting normalized CD spectra are shown in Figure~\ref{fig:CDRPE}(b). Remarkably, after normalization, all spectra from the purified nanobioconjugates collapsed onto a single curve across the entire spectral range, indicating that the CD response was characteristic of R-PE coupled to AuNP and was largely independent of the nanobioconjugate's composition. The key result of this work emerges from comparing these normalized spectra to those of R-PE in PBS, also normalized to unit absorption at 570\,nm. A nearly threefold increase in the CD amplitude at 570\,nm was observed for the AuNP/R-PE nanobioconjugates relative to free R-PE. This enhancement was observable already in the solution before purification (see Figure S2). The enhancement gradually decreased toward shorter wavelengths in the 500--570\,nm region, suggesting not only an amplitude enhancement but also a modification of the spectral lineshape induced by the coupling (see Figure S8). The CD enhancement was extracted from the dependence of the variations of the maximum CD near 570 nm as a function of the R-PE concentration obtained from Figure \ref{fig:assemblies}(d). The results are displayed in Figure~\ref{fig:CDRPE}(c). As expected, R-PE in PBS exhibited a linear dependence with a slope of 21\,mdeg$/\mu$M. A linear trend was also observed for the purified AuNP/R-PE nanobioconjugates, but with a significantly steeper slope of 62\,mdeg$/\mu$M. This corresponds to an enhancement factor of 2.9 at 570\,nm, consistent with the value derived from the normalized spectra. The enhancement value of 2.9 is a lower bound of the actual enhancement, owing to the absorption enhancement in R-PE (Figure S6). In contrast, the negative CD at the Trp-band was not amplified with respect to the average value of CD across the plateau in the 350-500 nm range (Figure S8). 

\subsection{Numerical analysis and comparison}
\begin{figure*}[!ht]
\centering\includegraphics[width=0.7\linewidth]{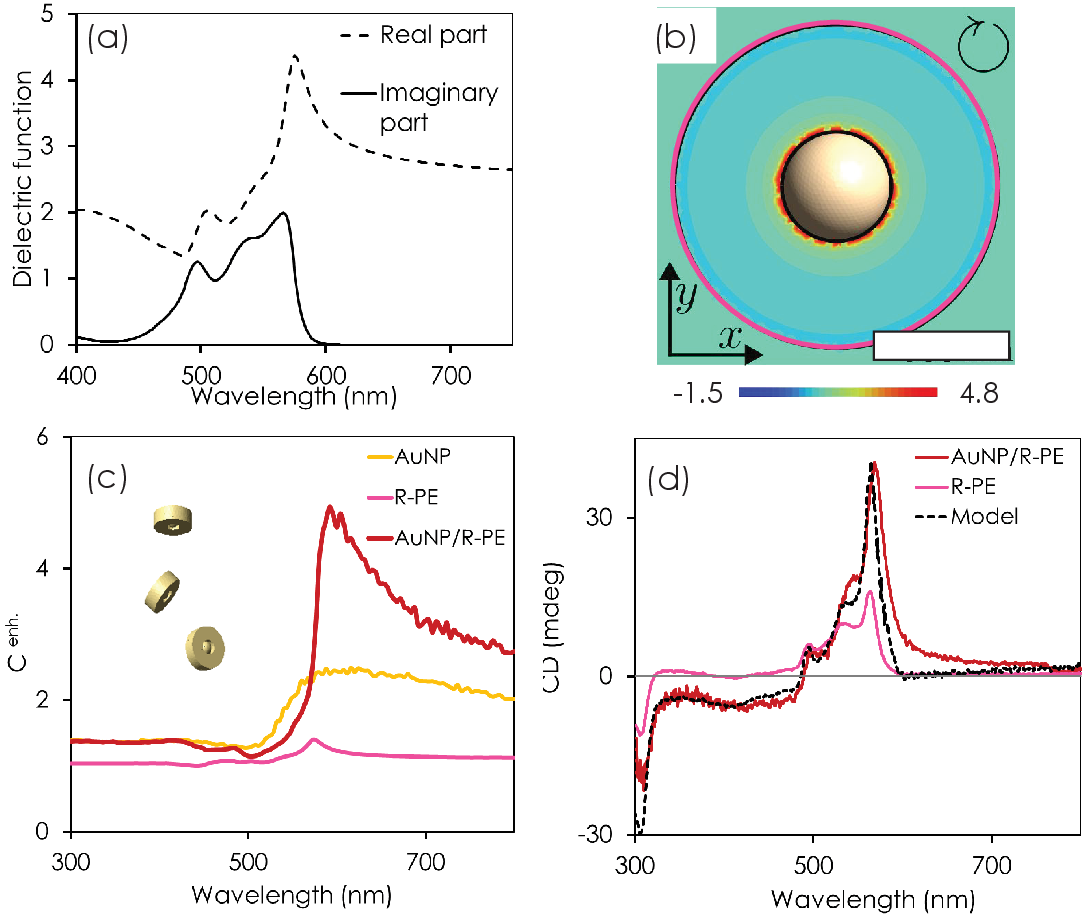}
\caption{Chirality density enhancement. (a) Complex dielectric permittivity of R-PE. (b) Calculated chirality-density map in a plane intersecting the center of the AuNP at its plasmon resonance (588 nm), for right-circularly polarized illumination. The pink contour indicates the geometrical extent of the R-PE shell. The scale bar is 5 nm. (c) Chirality density enhancement C$^{enh.}$ within the geometrical volume of the R-PE shell for three cases: R-PE in water (pink), an isolated AuNP in water (yellow), and the AuNP/R-PE nanobioconjugate (red). (d) Modeled CD spectrum of the AuNP/R-PE nanobioconjugate (dotted black line), obtained by combining the CD of R-PE in PBS (pink line) with the calculated chirality-density enhancement, absorption enhancement and a negative CD contribution associated with the AuNP. The experimentally measured CD of the AuNP/R-PE nanobioconjugate is shown for comparison (red line).}
\label{fig:Csimu}
\end{figure*}

The origin of both the CD enhancement and the modification of the spectral lineshape observed in the AuNP/R-PE nanobioconjugates, as compared to R-PE in PBS, was investigated through numerical simulations performed using a home-built electromagnetic code.\cite{Amboli_23} In the model, the AuNP/R-PE complex was represented by a 3\,nm-diameter gold nanosphere embedded at the center of a donut-shaped R-PE shell with an outer diameter of 12\,nm and a thickness of 3\,nm. \cite{Review_tunnel_cavity} The dielectric function of gold was taken from standard reference data,\cite{JC72} while that of R-PE, shown in Figure~\ref{fig:Csimu}(a), was extracted from the measured absorption spectrum of Figure~\ref{fig:concepts}(c). The surrounding medium was treated as homogeneous and dispersionless, with a refractive index of $n = 1.33$.

The electromagnetic response of the idealized nanobioconjugate was computed as a function of the wavelength in the 300-800 nm range using the Finite Element Method\cite{femcode,Amboli_23}. The solution AuNP/R-PE is modeled as a bi-periodic (along $x$ and $y$) arrangement of spheres and donuts with lattice size 30\,nm in a background medium of optical index 1.33. The periodic cell is made of a stack of 15 AuNP/R-PE dimers with centers originally centered in the cell in the $xy$ plane and spaced by 15\,nm along $z$. Each dimer is then randomly translated and rotated as shown in the inset of Figure~\ref{fig:assemblies}(e) to prevent from periodicity or orientation bias. The spectra are computed considering 10 random configurations with arbitrary translations and rotations for each dimer within the cell. The absorption spectra shown in Figure~\ref{fig:assemblies}(e) correspond to the absorption averaged over the 10 configurations. The incident field is a left or right circularly polarized plane wave with wavevector along $z$, \textit{i.e.} at normal incidence upon this periodic arrangement. The mesh size is set to $\approx\lambda/15$ in the background medium, 0.25\,nm in the gold sphere and 0.75\,nm in the surrounding donut. The computed near fields were subsequently used to compute the absorption normalized with respect to the incident power in the periodic cell as well as the optical chirality density within the donut-shaped volume surrounding the AuNP. 
Figure \ref{fig:Csimu}(b) presents the spatial distribution of the chirality density C calculated at $\lambda=588$ nm, in a plane in the middle of a single gold nanoparticle for circular incident polarization, normalized to the chirality density of the incident plane wave. Under linear polarization, localized regions of high values of $C$ were observed near the AuNP surface; however, the spatially averaged chirality density remained zero due to symmetry (see Figure S9). For circularly polarized excitation, the background value of $C$ matched that of the incident wave, while a clear enhancement of $C$ emerged in the close vicinity of the nanoparticle. Figure~\ref{fig:Csimu}(c) shows the maximum chirality density in this volume for circularly polarized illumination, normalized to the value for a plane wave in water: this will be noted C$^{enh.}$. For R-PE in water, C$^{enh.}$ exhibited only minor modulations around unity. For gold nanoparticles in water, C$^{enh.}$ increased gradually from unity below 500 nm to a plateau of $\approx$2.5 above 550 nm. In contrast, the AuNP/R-PE nanobioconjugate displayed a resonant C$^{enh.}$, peaking slightly above 5 near 588\,nm. This confirms that the elevated C$^{enh.}$ within the protein volume arises from the plasmonic resonance; without the AuNP, R-PE alone would not produce significant deviations from unity. The strong spectral dependence of the chirality density enhancement together with the absorption enhancement, suggest a corresponding modification of the CD spectrum.\cite{Alejandro_2018} 
The CD of the AuNP/R-PE nanobioconjugates, CD$^{AuNP/R-PE}$, was modeled as: 
\begin{equation}
    CD^{AuNP/R-PE}= (C_1.C^{enh.}+C_2.A^{RPE}).CD^{RPE}+ C_3.A^{AuNP}
\end{equation} 
\noindent where the three contributions were: the intrinsic CD of R-PE in PBS (CD$^{RPE}$) multiplied (i) by the CD enhancement C$^{enh.}$, scaled by factor of $C_1$ and (ii) by the absorption enhancement of the proteins A$^{RPE}$ (see Figure S6) scaled by a factor $C_2$; (iii) a plasmon-induced CD, proportional to the absorption of gold, modeled at $C_3$ per unit absorption of the AuNP (A$^{AuNP}$) normalized to unity at the LSPR. 
To test this hypothesis, Figure~\ref{fig:Csimu}(d) presents a modeled CD spectrum of the AuNP/R-PE nanobioconjugate (dotted black line), adjusted with $C_1$=1.4, $C_2$=1 and $C_2$=-4.5 mdeg, the latter being consistent with reported values for peptide-functionalized gold nanoparticles.\cite{Naik2011} 
The resulting spectrum displayed in Figure~\ref{fig:Csimu}(d) shows striking quantitative agreement with the experimentally measured CD of the AuNP/R-PE nanobioconjugates, both in overall amplitude and in the spectral evolution relative to R-PE in PBS. This agreement indicates that three mechanisms govern the CD response in the nanobioconjugates: (i) an enhancement of the molecular chiral response due to the increased local chirality density, (ii) an enhancement of the molecular absorption induced by the local intensity enhancement and (iii) a transfer of optical chirality from R-PE to the AuNP, contributing a negative CD near the plasmon resonance. The multiplicative factors ($C_1$, $C_2$) were obviously highly correlated and influenced by the AuNP’s position relative to R-PE, centered in our model (see Figure S10), and the CD baseline correction for colloidal AuNP suspensions (see Figure S7). The very good agreement between the spectral dependence of the modeled CD and measured CD for both the Trp-band near 300 nm and the Q-band in the 500-600 nm region modeled in Figure \ref{fig:Csimu}(d) underpin the CD enhancement demonstrated in Figures \ref{fig:CDRPE}(b-c).

\begin{figure*}[!ht]
    \centering
    \includegraphics[width=0.9\linewidth]{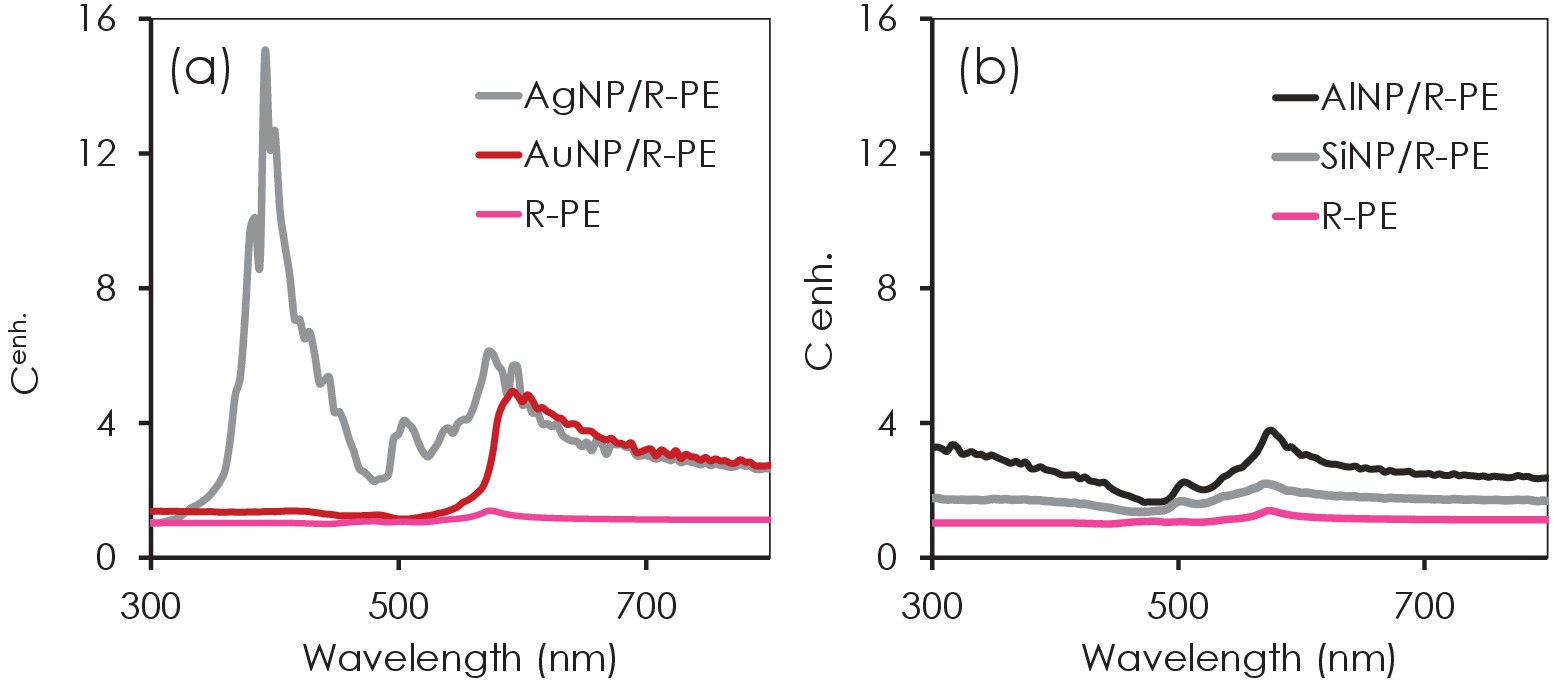}
    \caption{Design rules for CD enhancement. (a) Comparison of C$^{enh.}$ for nanobioconjugates of R-PE with silver and gold nanoparticles. (b) Comparison of C$^{enh.}$ for aluminum and silicon nanoparticles. C$^{enh.}$ in R-PE in PBS is recalled in each plot.}
    \label{fig:offresonance}
\end{figure*}

To extend our result to other systems and explore the relationship between C$^{enh.}$ in the near-field of nanoparticles and the enhancement of the CD response of proteins, we employed the same simulations to investigate the influence of the resonance coupling effect. Figure \ref{fig:offresonance}(a) presents the calculated C$^{enh.}$ for 3 nm gold and silver nanoparticles. For silver nanoparticles, a primary peak in C$^{enh.}$ was observed at the LSPR of silver with a 15-fold enhancement. Additionally, a second peak of C$^{enh.}$ emerged near the Q-band of R-PE with values comparable to those observed for gold. Figure \ref{fig:offresonance}(b) illustrates the case of 3 nm  aluminum and silicon nanoparticles. With aluminum nanoparticles, C$^{enh.}$ showed a modest increase near the Q-band of R-PE, peaking at 3.6. For Si nanoparticles, the peak values never exceeded 2.2. In all cases, the electronic transitions of R-PE at the Q-band induced an increase of C$^{enh.}$. 

The peak values reached depended strongly on the material used for the nanoparticles. All materials were absorbing in the spectral region investigated. However, it can be noted that the values of C$^{enh.}$ were closely related to the origin of the absorption in the nanoparticles. Indeed, absorption originating from bound electrons was associated with small values of C$^{enh.}$. This was the case for nanoparticles made of Si, for gold below 500 nm and to some extent for Al which exhibits interband transitions centered near 800 nm, tailing down to the visible part of the spectrum.\cite{Rakic95} In contrast, absorption associated with free electrons induced large values of  C$^{enh.}$ both at the LSPR of the nanoparticles and at the electronic transitions of the analyte. 

\section{DISCUSSION}
Electrostatic assembling of R-phycoerythrin (R-PE) with gold nanoparticles (AuNP) was achieved, followed by purification to isolate the hybrid nanobioconjugates. The relative concentrations of R-PE and AuNP were determined via UV-Vis absorption spectroscopy. While the spectral signatures of both R-PE and AuNP were clearly observed, the measured absorption spectrum could not be decomposed as a simple linear combination of the individual spectra of bare AuNP and R-PE. This observation was further complicated by the expected redshift of the AuNP's LSPR upon transitioning from an aqueous environment to the R-PE matrix.

Numerical simulations were employed to disentangle the contributions of R-PE and AuNPs to the total absorption. These analyses revealed a pronounced non-local response from both components, which significantly complicated the interpretation of the absorption spectra. However, the close agreement between experimental and simulated absorption spectra (Figure~\ref{fig:assemblies}(d)) supports the hypothesis that the purification process preserved the electromagnetic interactions between AuNP and R-PE across all sample compositions. The R-PE-to-AuNP ratio increased with the initial composition of the solutions, although it never reached a 1:1 stoichiometry. Transmission electron microscopy (TEM) images confirmed the presence of large protein aggregates with AuNP distributed among the proteins. 

Circular dichroism (CD) measurements have shown a partial alteration of the R-PE's secondary structure within the AuNP/R-PE nanobioconjugates. Despite the strong UV absorption of gold near 200~nm, which limited the accuracy of structural determination, the characteristic $\alpha$-helix signature of R-PE loses its contribution at 210 nm, corresponding to the $\pi$-$\pi$* transition while the n-$\pi$* at 220 nm is preserved across all compositions (Figure~\ref{fig:assemblies}(c) and Figure S4).\cite{Miles2016} The decreased values near 210 nm have been observed when R-PE was associated with nanoparticles.\cite{Kovalev2013} However, in this case, the CD in the Q-band was also reduced in stark contrast with the enhancement we have observed in Figures \ref{fig:CDRPE}(b-c). The CD signatures of the Trp-band (300~nm) was also preserved. The preservation of the Trp-band was also a good indication that the secondary structure of R-PE was not heavily affected in the nanobioconjugates.\cite{Kovalev2013} After normalization to the composition extracted from UV-Vis analysis, all CD spectra exhibited excellent overlap, validating the reproducibility of our analytical approach.

Most strikingly, the CD signal in the Q-band (500-600~nm) exhibited a 3-fold enhancement in the nanobioconjugates (62~mdeg/$\mu$M) compared to R-PE in phosphate-buffered saline (21~mdeg/$\mu$M) while it remained mostly unchanged in the Trp-band. This enhancement was much stronger than the molecular absorption enhancement (Figure S6).\cite{Naik2010} It was observed across all compositions, with notable changes in both the relative amplitudes and absolute intensities of the three Q-bands  (Figure~\ref{fig:assemblies}(b)) without inversion in the CD sign.\cite{LinkDNA2019} The agreement between calculated and measured absorption spectra, coupled with the global conservation of secondary and tertiary structures in the CD spectra, emphasizes the strong electromagnetic coupling between R-PE and AuNP in the nanobioconjugates at the LSPR of the AuNP. 

Several mechanisms have been proposed to explain CD enhancement in plasmonic-chiral analyte systems \cite{Naik2010, Weiss2022}. In solution, care must be taken to avoid artifacts from chiral plasmonic heterostructures induced by chiral adsorbates \cite{Link2019, Parola_2022, Marzan_2020}. However, TEM images of our nanobioconjugates (Figure~\ref{fig:assemblies}(b)) did not reveal such heterostructures, ruling out this effect. It is also admitted that the chirality density around a plasmonic resonator should average to zero, but this should not be the case with the ring shaped R-PE used in our work (see Figure S11).\cite{Dionne2013} 

Plasmonic enhancement of chiral optical responses is typically attributed to the interaction between the electromagnetic near-field of the nanoparticle and the chiral analyte \cite{Weiss_ACS_2016, DiskGovorov_2013, Oleg_NanoLett_2013, García-Guirado2020,LinkDNA2019,Naik2010}. In our system, the electronic transitions of R-PE associated with the Q-band are resonantly coupled with the plasmon mode of the AuNPs. While mode hybridization has been proposed under such conditions \cite{Alejandro_2018,LinkDNA2019}, no spectral splitting or sign inversion was observed in our measurements (Figure~\ref{fig:CDRPE}). 

Numerical simulations revealed a up to 5-fold enhancement of the chirality density localized near the AuNP surface and a 40$\%$ increase of absorption in R-PE, with a spectral dependence modified by the presence of R-PE (Figure~\ref{fig:Csimu}(c,d) and Figure S6). This enhancement decays rapidly with the distance from the AuNP, which is consistent with near-field spatial distributions (Figure~\ref{fig:Csimu}(b)). Surprisingly, our calculations suggest that all chromophores in the R-PE protein were equally affected by the chirality density enhancement (Figure~\ref{fig:Csimu}(d)), despite the expected localization of near-field effects. Transient grating and photon-echo spectroscopies have demonstrated ultrafast exciton delocalization across chromophore pairs in light-harvesting complexes, with sub-100-fs self-trapping and ~1-ps energy transport dynamics \cite{JPhysChem_Andrew_2011, Deloc_exc_NatComm_2014}. These findings align with theories of delocalized excitons, where electronic wavefunctions span molecular aggregates to enhance energy-funneling efficiency \cite{RevModPhys_2018}. Our results (Figure~\ref{fig:Csimu}(d)) suggest that such delocalization may also manifest in the CD spectra of phycobiliproteins, providing a potential explanation for the uniform enhancement observed across all chromophores. We must keep in mind that the magnitudes of the calculated C$^{enh.}$ and A$^{RPE}$ relied on different parameters like: the actual location of the AuNP with respect to the R-PE structure, the optical constants of R-PE which contains the static dielectric constant (arbitrarily set to 1.6 here), the extent of the delocalization of the wavefunctions in R-PE, etc. Although the enhancement of CD determined experimentally was accurate (Figure~\ref{fig:CDRPE}(b)) within the composition uncertainties, the magnitudes of the numerical C$^{enh.}$ and A$^{RPE}$ should be taken more cautiously. In contrast, the spectral lineshapes of the calculated quantities were most likely accurate and explained the spectral variation of the measured CD. Our results suggest that the effects described here should be even more pronounced in the case of chromophores small enough, i.e. 1-2 nm, to fully experience the enhancement. 

To assess the generalization of our findings, we calculated C$^{enh.}$ for R-PE nanobioconjugates with silver, aluminum, and silicon nanoparticles (Figure~\ref{fig:offresonance}). Only silver and gold exhibited significant enhancement at the R-PE Q-band and up to 15-fold in the AgNP/R-PE system at the LSPR. In contrast, silicon and aluminum showed minimal modulation. This trend is consistent with the resonant coupling of pure free-electron oscillations in noble metals to the electronic transitions of the chiral analyte, implying that our observations may extend to other systems where the LSPR, and more generally free-electron oscillations, overlaps with the analyte's electronic transitions.

\section{CONCLUSION}
In this study, we demonstrated that plasmonic nanoparticles can significantly amplify the circular dichroism (CD) response of biological chromophores by engineering superchiral near fields. Through a combination of experimental measurements and numerical simulations, we elucidated the origin of the CD enhancement and modification of the spectral lineshape in AuNP/R-PE nanobioconjugates. Our results show that chirality density enhancement (C$^{enh.}$) and absorption enhancement (A$^{RPE}$), driven by resonant coupling between the free electrons oscillations of noble metal nanoparticles and the electronic transitions of analytes, play a key role in amplifying the chiral response of adjacent biomolecules. 
The modeled CD spectra, which were constructed by combining the intrinsic CD of R-PE, the calculated C$^{enh.}$ and A$^{RPE}$, and a possible negative contribution from p)lasmon-induced CD, exhibited excellent quantitative agreement with the experimental measurements. This agreement validates our numerical approach and highlights the dual mechanisms governing the CD response in these nanobioconjugates: (1) enhancement of the chiral molecular response due to increased local chirality density and absorption and (2) transfer of optical chirality from biomolecules to nanoparticles. 
These findings provide critical insights into designing plasmonic architectures for chiral sensing and asymmetric photochemistry. Tailoring the material, size, and spatial arrangement of nanoparticles may allow for further optimization of response enhancement and greater sensitivity in chiral detection. Future work could extend these principles to other biomolecular systems or develop metasurfaces and nanodevices that exploit superchiral fields for biosensing, enantiomeric separation and chiral catalysis applications.

\section{MATERIAL and METHODS}
\noindent\textit{Materials.} Hydrogen tetrachloroaurate(III) hydrate (HAuCl$_4$.3H$_2$O, Sigma-Aldrich, 50$\%$ Au basis), 1,2,3,4-tetrahydronaphthalene (tetraline for synthesis, $>$98$\%$, Sigma), oleylamine (OLA, Acros, 80-90$\%$), tert-butylamine borane (97$\%$, Alfa aesar), acetylthiocholine chloride (>99$\%$, sigma), deuterium oxide (99.9$\%$, Eurisotop), diethyl ether (99$\%$, prolabo), HCl (37$\%$, sigma), sodium phosphate ($\geq$99$\%$, sigma), sodium chloride ($>$99.5$\%$, sigma), sodium hydroxide (NaOH, sigma) were used as received. Absolute ethanol (abs. EtOH, normapur), methanol (MeOH, normapur), toluene (rectapur), acetone (rectapur) were purchased from VWR. R-phycoerythrin (R-PE) from red alguae was bought from antibodies.com.\\
The initial solution of R-PE ($\geq$10.0 mg/mL in 150 mM sodium phosphate, 60$\%$ ammonium sulfate, 1 mM EDTA, 1 mM sodium azide, pH 7.0) was dialyzed against 0.1M sodium phosphate and 0.1M sodium chloride and the pH was adjusted at pH=7.2 with NaOH 1M. MilliQ water (18.3 M$\Omega$.cm) was used in all experiments.\\
Glassware was cleaned with aqua regia, then repeatedly rinsed with demineralized water, and left in water bath overnight. \noindent\textit{Caution:} Aqua regia is highly toxic and corrosive; it requires personal protective equipment and should be handled under a fume hood.\\
\noindent\textit{3.6 nm gold NPs synthesis.} The synthesis was adapted from Ref. \cite{AuNP_Peng}. Briefly, in a 20mL vial in an ice water bath at 10°C, 50mg of HAuCl$_4$.3H$_2$O (0.127mmol) were dissolved in 5mL of tetralin and 5mL OLA under stirring. A solution containing 24mg (0.25mmol) of tert-butylamine borane, 0.5mL of tetralin and 0.5mL of OLA was mixed by sonication until full dissolution and quickly injected into the precursor solution. The solution changed to a deep purple color within 5min. The mixture was allowed to react at 10°C for 1h. Then, acetone was added to precipitate the AuNP, that were collected by centrifugation in 1 centrifuge tube (6000rpm, 3min), and redispersed in toluene. The AuNP were precipitated a second time with a mixture of acetone and MeOH, centrifugated (6000rpm, 3min), and redispersed in 5mL toluene.\\
\noindent\textit{Deprotection of acetylthiocholine chloride.} The synthesis was adapted from Ref. \cite{Tibaldi2019}. In a 25mL round bottom flask with a condenser and a N$_2$ atmosphere, acetylthiocholine chloride (108 mg, 0.54mmol) was dissolved in 1mL 12M HCl and 4mL abs. EtOH. The stirred solution was heated to reflux overnight. The reaction was concentrated under reduced pressure and dried under high vacuum. The residue was washed twice with abs. EtOH and then concentrated to 1mL under vacuum. Diethyl ether (10mL) was added and a white precipitate appeared, isolated by centrifugation. The supernatant was then removed and diethyl ether was added to wash the precipitate. The solid was recovered after centrifugation and dried under vacuum, providing (2-mercaptoethyl-N,N,N-trimethylammonium chloride) as a white precipitate (70mg, 90$\%$). The purity of the precipitate was confirmed by $^1$H and $^{13}$C NMR (400 MHz), D$_2$O. This material was solubilized in 1.5mL abs. EtOH (0.3M). \\
\noindent\textit{Ligand exchange from OLA to 2-mercaptoethyl-N,N,N-trimethylammonium (meta).} In a centrifuge tube, containing 5ml of AuNP in toluene, 1mL of a solution of 2-mercaptoethyl-N,N,N-trimethylammonium chloride at 0.3M in abs EtOH was added. A blue precipitate appeared that was isolated after vigorous mixing, sonication, and centrifugation (6000rpm, 3min). Then, the precipitate was washed with the ligand solution and abs. EtOH, followed by sonication and centrifugation (6000rpm, 3min). Finally, the precipitate was redispersed in 5mL milliQ water. The AuNP-meta (AuNP in the main text) dispersion was analyzed by UV vis (5µL in 1mL, lplasmon=511nm, Aplasmon=0.29, A450=0.28, [NP]=23µM) and by TEM (<D>=3,5nm, $\sigma$=0,34nm).\\
\noindent\textit{AuNP / R-PE nanobioconjugate.} In a 20mL vial, containing 3mL of milliQ water, 92µL of 23µM AuNP was added while the dispersion is mixed on an orbital mixer. Then, a volume (11, 17, 22, 27, 41, 55µL) of a solution of R-PE at 44µM was added. The reaction was pursued for 1h. The dispersion was centrifugated (11000 rpm, 5min), the supernatant was removed and the solid was redispersed in 3mL of milliQ water. \\

\noindent\textit{Characterizations.} The absorbance A of the AuNP and R-PE solutions were measured in the 300-800 nm range on a CARY 500 in a quartz cuvette with a light path of 1 cm. The extinction Ext was determined as Ext=-log(T). The CD spectra were also measured on a J815 JASCO circular dichroism spectrometer in a quartz cuvette with a 0.2 cm light path. 
The CD was corrected by subtracting the CD measured on a cuvette with the same light path containing pure water. The contribution of the absorption of the R-PE to the AuNP/R-PE solution was determined using a peak decomposition. The AuNP absorption was modeled using 3 Gaussian peaks, two for the interband transition below 500 nm and one around 550 nm for the plasmon resonance. The amplitudes, broadening and position of the peaks were adjusted. The R-PE absorption was modeled using the measured R-PE absorption in solution, rescaled by a multiplicative factor so as to minimize the difference between the measured and modeled absorption. The diameter of the AuNP was determined using a JEOL transmission electron microscope working at an accelerating voltage of 200 kV. A droplet of AuNP was deposited on a carbon suspended film and was left to dry before introduction in the microscope. \\

\noindent\textit{Calculations. }  The numerical results rely on open source models\cite{gmsh,getdp,femcode} based on Finite Elements \cite{Demesy10,Amboli_23}.

\section{Associated content}

Supplementary Information content: charge state of the different species through zeta-potential measurements; transmission and CD measurements for the evidence of the effect of electrostatic assembling; transmission measurements on initial solutions, supernatants and redispersions; CD-DUV spectra of all nanobioconjugates; decomposition of the absorption spectra and the numerical explanation of the differences observed between models and experiments; raw CD spectra of baselines and nanobioconjugates; different normalizations of CD to allow for a better comparison between nanobioconjugates and initial R-PE; calculated intensity and chirality density maps around a nanoparticle for different incident polarizations, variations of the calculated C$^{enh.}$ for different AuNP/R-PE orientations and different cuts of the chirality density map.

\section*{Funding}
This work has been carried out thanks to the support of the French National Research Agency (ANR) under grant n$^{\circ}$ ANR-23-CE42-0019 and is part of the BERNARDO project of PEPR LUMA as part of the France 2030 program under grant ANR-24-EXLU-0001. 

\section{Author contributions}
B.G., Y.P conceived the ideas and directed the project. V.H and M.V. synthesized the samples under the direction of Y.P. and analyzed the chemistry data. S.B. and M.S. supervised the chemistry characterizations and provided support in understanding the properties of the proteins. G.D. and N.B. performed the numerical simulations and provided support in the analysis of the experimental data performed by M.-S.P., M.M. and B.G.. B.G. and Y.P. wrote the paper.   

\newpage
\bibliography{main_bibliography}

@article{García-Guirado2020,
  title={Enhanced chiral sensing with dielectric nanoresonators},
  author={Garcia-Guirado, Jose and Svedendahl, Mikael and Puigdollers, Joaquim and Quidant, Romain},
  journal={Nano Letters},
  volume={20},
  number={1},
  pages={585--591},
  year={2019},
  publisher={ACS Publications}
}

@article{Nesterov2016,
  title={The role of plasmon-generated near fields for enhanced circular dichroism spectroscopy},
  author={Nesterov, Maxim L and Yin, Xinghui and Sch{\"a}ferling, Martin and Giessen, Harald and Weiss, Thomas},
  journal={ACS Photonics},
  volume={3},
  number={4},
  pages={578--583},
  year={2016},
  publisher={ACS Publications}
}

@article{Adhikari2022,
  title={Optically probing the chirality of single plasmonic nanostructures and of single molecules: Potential and obstacles},
  author={Adhikari, Subhasis and Orrit, Michel},
  journal={ACS Photonics},
  volume={9},
  number={11},
  pages={3486--3497},
  year={2022},
  publisher={ACS Publications}
}

@article{Arteaga2016,
  title={Relation between 2D/3D chirality and the appearance of chiroptical effects in real nanostructures},
  author={Arteaga, Oriol and Sancho-Parramon, Jordi and Nichols, Shane and Maoz, Ben M and Canillas, Adolf and Bosch, Salvador and Markovich, Gil and Kahr, Bart},
  journal={Opt. Express},
  volume={24},
  number={3},
  pages={2242--2252},
  year={2016},
  doi={10.1364/OE.24.002242},
  publisher={Optical Society of America}
}

@article{OMarkovic2025,
  title={From Ideal to Real: Chiroptical Consequences of Asymmetries in Plasmonic Nano-Slit Metasurfaces},
  author={Markovic, Obren and Chaabani, Wajdi and Schwob, Catherine and Mivelle, Mathieu and Gallas, Bruno},
  journal={ACS Photonics},
  volume={12},
  number={11},
  pages={6360–6367},
  year={2025},
  publisher={American Chemical Society},
  doi = {10.1021/acsphotonics.5c01823}
}

@article{McPeak2019,
  title={Correlation of circular differential optical absorption with geometric chirality in plasmonic meta-atoms},
  author={Wilson, Jon C and Gutsche, Philipp and Herrmann, Sven and Burger, Sven and McPeak, Kevin M},
  journal={Optics Express},
  volume={27},
  number={4},
  pages={5097--5115},
  year={2019},
  doi={10.1364/OE.27.005097},
  publisher={Optical Society of America}
}

@article{Nicolas2023,
  title={True circular dichroism in optically active achiral metasurfaces and its relation to chiral near-fields},
  author={Nicolas, Mathieu and Walmsness, Per Magnus and Amboli, Jayeeta and Zhang, Lu and Demesy, Guillaume and Bonod, Nicolas and Boujday, Souhir and Kildemo, Morten and Gallas, Bruno},
  journal={ACS Appl. Opt. Mater.},
  volume={1},
  number={8},
  pages={1360--1366},
  year={2023},
  publisher={ACS Publications}
}

@article{Cui2023,
  title={Achiral magnetic photonic antenna as a tunable nanosource of chiral light},
  author={Cui, Lingfei and Yang, Xingyu and Reynier, Beno{\^\i}t and Schwob, Catherine and Bidault, S{\'e}bastien and Gallas, Bruno and Mivelle, Mathieu},
  journal={ACS Photonics},
  volume={10},
  number={11},
  pages={3850--3857},
  year={2023},
  publisher={ACS Publications}
}

@article{Weiss2022,
  title={Nanophotonic chiral sensing: how does it actually work?},
  author={Both, Steffen and Sch{\"a}ferling, Martin and Sterl, Florian and Muljarov, Egor A and Giessen, Harald and Weiss, Thomas},
  journal={ACS Nano},
  volume={16},
  number={2},
  pages={2822--2832},
  year={2022},
  doi={10.1021/acsnano.1c09796},
  publisher={ACS Publications}
}

@article{JC72,
  title={Optical constants of the noble metals},
  author={Johnson, Peter B and Christy, RW},
  journal={Phys. Rev. B},
  volume={6},
  number={12},
  pages={4370},
  year={1972},
  publisher={APS}
}

@article{Rho2020,
  title={Electromagnetic chirality: from fundamentals to nontraditional chiroptical phenomena},
  author={Mun, Jungho and Kim, Minkyung and Yang, Younghwan and Badloe, Trevon and Ni, Jincheng and Chen, Yang and Qiu, Cheng-Wei and Rho, Junsuk},
  journal={Light: Science \& Applications},
  volume={9},
  number={1},
  pages={139},
  year={2020},
  url={https://www.nature.com/articles/s41377-020-00367-8},
  publisher={Nature Publishing Group UK London}
}

@article{TangCohen2010,
  title={Optical chirality and its interaction with matter},
  author={Tang, Yiqiao and Cohen, Adam E},
  journal={Phys. Rev. Lett.},
  volume={104},
  number={16},
  pages={163901},
  year={2010},
  doi={10.1103/PhysRevLett.104.163901},
  publisher={APS}
}

@article{Giessen2012,
  title={Formation of chiral fields in a symmetric environment},
  author={Sch{\"a}ferling, Martin and Yin, Xinghui and Giessen, Harald},
  journal={Opt. Express},
  volume={20},
  number={24},
  pages={26326--26336},
  year={2012},
  doi={10.1364/OE.20.026326},
  publisher={Optical Society of America}
}

@article{Genet2014,
  title={Chiral near fields generated from plasmonic optical lattices},
  author={Canaguier-Durand, Antoine and Genet, Cyriaque},
  journal={Physical Review A},
  volume={90},
  number={2},
  pages={023842},
  year={2014},
  doi={10.1103/PhysRevA.90.023842},
  publisher={APS}
}

@article{Link2019,
  title={Unraveling the origin of chirality from plasmonic nanoparticle-protein complexes},
  author={Zhang, Qingfeng and Hernandez, Taylor and Smith, Kyle W and Hosseini Jebeli, Seyyed Ali and Dai, Alan X and Warning, Lauren and Baiyasi, Rashad and McCarthy, Lauren A and Guo, Hua and Chen, Dong-Hua and others},
  journal={Science},
  volume={365},
  number={6460},
  pages={1475-1478},
  year={2019},
  doi={10.1126/science.aax5415},
  publisher={American Association for the Advancement of Science}
}

@article{Hendry2010,
  title={Ultrasensitive detection and characterization of biomolecules using superchiral fields},
  author={Hendry, Euan and Carpy, T and Johnston, J and Popland, M and Mikhaylovskiy, RV and Lapthorn, AJ and Kelly, SM and Barron, LD and Gadegaard, N and Kadodwala, MJNN},
  journal={Nat. Nanotechnol.},
  volume={5},
  number={11},
  pages={783--787},
  year={2010},
  url={https://www.nature.com/articles/nnano.2010.209},
  publisher={Nature Publishing Group UK London}
}

@article{Naik2010,
  title={Theory of circular dichroism of nanomaterials comprising chiral molecules and nanocrystals: plasmon enhancement, dipole interactions, and dielectric effects},
  author={Govorov, Alexander O and Fan, Zhiyuan and Hernandez, Pedro and Slocik, Joseph M and Naik, Rajesh R},
  journal={Nano Lett.},
  volume={10},
  number={4},
  pages={1374-1382},
  year={2010},
  publisher={ACS Publications},
  doi={10.1021/nl100010v}
}

@article{Kumar2019,
  title={Recent Advances in Chiral Plasmonics - Towards Biomedical Applications},
  author={Kumar, Jatish and Liz-Marz\'{a}n, Luis M.},
  journal={Bull. Chem. Soc. Jpn.},
  volume={92},
  number={-},
  pages={30--37},
  year={2019},
  doi={10.1246/bcsj.20180236},
  publisher={Society for Applied Spectroscopy}
}

@article{Alu2017,
  title={Chirality detection of enantiomers using twisted optical metamaterials},
  author={Zhao, Yang and Askarpour, Amir N. and Sun, Liuyang and Shi, Jinwei and Li, Xiaoqin and Al\'{u}, Andrea},
  journal={Nat. Comm.},
  volume={8},
  number={-},
  pages={14180},
  year={2017},
  publisher={Nature},
  url = {https://www.nature.com/articles/ncomms14180}
}

@article{Hendry2013,
  title = {Superchiral electromagnetic fields created by surface plasmons in nonchiral metallic nanostructures},
  author = {Davis, T. J. and Hendry, E.},
  journal = {Phys. Rev. B},
  volume = {87},
  issue = {8},
  pages = {085405},
  numpages = {5},
  year = {2013},
  doi={10.1103/PhysRevB.87.085405},
  publisher = {American Physical Society},
}

@article{Naik2011,
  title={Plasmonic Circular Dichroism of Peptide-Functionalized Gold Nanoparticles},
  author={Slocik, Joseph M. and Govorov and Naik, Rajesh R},
  journal={Nano Lett.},
  volume={11},
  number={2},
  pages={701-705},
  year={2011},
  doi={10.1021/nl1038242},
  publisher={ACS Publications}
}

@Article{C1JM12345A,
author ={Govorov, Alexander O. and Gun'ko, Yurii K. and Slocik, Joseph M. and Gérard, Valérie A. and Fan, Zhiyuan and Naik, Rajesh R.},
title  ={Chiral nanoparticle assemblies: circular dichroism, plasmonic interactions, and exciton effects},
journal  ={J. Mater. Chem.},
year  ={2011},
volume  ={21},
issue  ={42},
pages  ={16806-16818},
publisher  ={The Royal Society of Chemistry},
doi  ={10.1039/C1JM12345A},
}

@Article{Gunko2007,
author ={Mícheál P. Moloney and Yurii K. Gun'ko and John M. Kelly},
title  ={Chiral highly luminescent CdS quantum dots},
journal  ={Chem. Commun.},
year  ={2007},
volume  ={38},
issue  ={14},
pages  ={3900-3902},
publisher  ={The Royal Society of Chemistry},
doi  ={10.1039/B704636G},
}

@article{Davis2012,
  title={Chiral Electromagnetic Fields Generated by Arrays of Nanoslits},
  author={Hendry, E. and Mikhaylovskiy, R.V. and Barron, L.D. and Kadodwala, M. and Davis, J.},
  journal={Nano Lett.},
  volume={12},
  number={7},
  pages={3640--3644},
  year={2012},
  doi={10.1021/nl3012787},
  publisher={ACS Publications}
}

@article{Nicolas2025,
  title = {Probing peptide adsorption kinetics and regioselectivity via multipolar plasmonic modes of gold resonators},
  author = {Nicolas, Mathieu and Yang, Shuhui and Méthivier, Christophe and Boujday, Souhir and Gallas, Bruno},
  journal = {Appl. Phys. B},
  volume = {131},
  issue = {-},
  pages = {162},
  numpages = {-},
  year = {2025},
  month = {Jul},
  publisher = {Springer Nature},
  doi = {10.1007/s00340-025-08525-9},
}

@article{Alejandro_2018,
  title = {Optical Chirality in Dispersive and Lossy Media},
  author = {V\'azquez-Lozano, J. Enrique and Mart\'{\i}nez, Alejandro},
  journal = {Phys. Rev. Lett.},
  volume = {121},
  issue = {4},
  pages = {043901},
  numpages = {7},
  year = {2018},
  month = {Jul},
  publisher = {American Physical Society},
  doi = {10.1103/PhysRevLett.121.043901},
}

@article{Oleg_NanoLett_2013,
author = {Lu, Fang and Tian, Ye and Liu, Mingzhao and Su, Dong and Zhang, Hui and Govorov, Alexander O. and Gang, Oleg},
title = {Discrete Nanocubes as Plasmonic Reporters of Molecular Chirality},
journal = {Nano Letters},
volume = {13},
number = {7},
pages = {3145-3151},
year = {2013},
doi = {10.1021/nl401107g},
}

@article{Multipole_PRB2008,
  title = {Optical properties of coupled metal-semiconductor and metal-molecule nanocrystal complexes: Role of multipole effects},
  author = {Yan, Jie-Yun and Zhang, Wei and Duan, Suqing and Zhao, Xian-Geng and Govorov, Alexander O.},
  journal = {Phys. Rev. B},
  volume = {77},
  issue = {16},
  pages = {165301},
  numpages = {9},
  year = {2008},
  month = {Apr},
  publisher = {American Physical Society},
  doi = {10.1103/PhysRevB.77.165301}
}

@article{DiskGovorov_2013,
author = {Maoz, Ben M. and Chaikin, Yulia and Tesler, Alexander B. and Bar Elli, Omri and Fan, Zhiyuan and Govorov, Alexander O. and Markovich, Gil},
title = {Amplification of Chiroptical Activity of Chiral Biomolecules by Surface Plasmons},
journal = {Nano Letters},
volume = {13},
number = {3},
pages = {1203-1209},
year = {2013},
doi = {10.1021/nl304638a},
}

@article{Weiss_ACS_2016,
author = {Nesterov, Maxim L. and Yin, Xinghui and Schäferling, Martin and Giessen, Harald and Weiss, Thomas},
title = {The Role of Plasmon-Generated Near Fields for Enhanced Circular Dichroism Spectroscopy},
journal = {ACS Photonics},
volume = {3},
number = {4},
pages = {578-583},
year = {2016},
doi = {10.1021/acsphotonics.5b00637}
}

@article{Parola_2022,
author = {Carone, Antonio and Mariani, Pablo and D{\'e}sert, Anthony and Romanelli, Marco and Marcheselli, Jacopo and Garavelli, Marco and Corni, Stefano and Rivalta, Ivan and Parola, Stephane},
title = {Insight on Chirality Encoding from Small Thiolated Molecule to Plasmonic Au@Ag and Au@Au Nanoparticles},
journal = {ACS Nano},
volume = {16},
number = {1},
pages = {1089-1101},
year = {2022},
doi = {10.1021/acsnano.1c08824},
}

@article{Marzan_2020,
author = {Severoni, Emilia and Maniappan, Sonia and Liz-Marzán, Luis M. and Kumar, Jatish and García de Abajo, F. Javier and Galantini, Luciano},
title = {Plasmon-Enhanced Optical Chirality through Hotspot Formation in Surfactant-Directed Self-Assembly of Gold Nanorods},
journal = {ACS Nano},
volume = {14},
number = {12},
pages = {16712-16722},
year = {2020},
doi = {10.1021/acsnano.0c03997},
}

@article{Amboli_23,
author = {Jayeeta Amboli and Bruno Gallas and Guillaume Dem\'{e}sy and Nicolas Bonod},
journal = {Opt. Express},
number = {26},
pages = {43147-43162},
publisher = {Optica Publishing Group},
title = {Design and analysis of chiral and achiral metasurfaces with the finite element method},
volume = {31},
month = {Dec},
year = {2023},
doi = {10.1364/OE.500540},
}

@article{Bunster2004,
author = {Martínez-Oyanedel, J. and Contreras-Martel, C. and Bruna, C. and Bunster M.},
title = {Structural-functional analysis of the oligomeric protein R-phycoerythrin},
journal = {Biol Res.},
volume = {37},
number = {4},
pages = {733--745},
year = {2004},
doi = {10.4067/s0716-97602004000500003},
}

@article{Schreiber2013_DNA_Nanostructures_CD,
  author = {Schreiber, R. and Luong, N. and Fan, Z. and Kuzyk, A. and Nickels, P.C. and Zhang, T.},
  title = {Chiral plasmonic DNA nanostructures with switchable circular dichroism},
  journal = {Nature Communications},
  year = {2013},
  volume = {4},
  pages = {2948},
  doi = {10.1038/ncomms3948}
}

@article{PaivaMarques2020_ChiralPlasmonics_Sensors,
  author = {Paiva-Marques, W. and Reyes Gómez, F. and Oliveira, O., Jr. and Mejía-Salazar, J.R.},
  title = {Chiral Plasmonics and Their Potential for Point-of-Care Biosensing Applications},
  journal = {Sensors},
  year = {2020},
  volume = {20},
  number = {3},
  pages = {944},
  doi = {10.3390/s20030944}
}

@article{JPhysChem_Andrew_2011,
author = {Womick, Jordan M. and Liu, Haoming and Moran, Andrew M.},
title = {Exciton Delocalization and Energy Transport Mechanisms in R-Phycoerythrin},
journal = {The Journal of Physical Chemistry A},
volume = {115},
number = {12},
pages = {2471-2482},
year = {2011},
doi = {10.1021/jp111720a},
}

@article{RevModPhys_2018,
  title = {Delocalized excitons in natural light-harvesting complexes},
  author = {Jang, Seogjoo J. and Mennucci, Benedetta},
  journal = {Rev. Mod. Phys.},
  volume = {90},
  issue = {3},
  pages = {035003},
  numpages = {49},
  year = {2018},
  month = {Aug},
  publisher = {American Physical Society},
  doi = {10.1103/RevModPhys.90.035003},
}

@article{Deloc_exc_NatComm_2014,
  title = {Dynamic localization of electronic excitation in photosynthetic complexes revealed with chiral two-dimensional spectroscopy},
  author = {Fidler, A. and Singh, V. and Long, P. and  Peter, D. Dahlberg and Gregory, S. Engel},
  journal = {Nat. Commun.},
  volume = {5},
  issue = {-},
  pages = {3286},
  numpages = {-},
  year = {2014},
  month = {-},
  publisher = {Nature},
  doi = {10.1038/ncomms4286},
}

@article{Review_algae,
  title = {Marine algae colorants: Antioxidant, anti-diabetic properties and applications in food industry},
  author = {Temjensangba Imchen, Keisham Sarjit Singh},
  journal = {Algal Research},
  volume = {69},
  issue = {-},
  pages = {102898},
  numpages = {14},
  year = {2023},
  month = {-},
  publisher = {Elsevier},
  doi = {10.1016/j.algal.2022.102898},
}

@article{Review_tunnel_cavity,
  title = {Properties and potential applications of bioconjugates of R-phycoerythrin with Ag$^o$ or CdS nanoparticle synthesized in its tunnel cavity: A review},
  author = {Olga Bekasova},
  journal = {International Journal of Biological Macromolecules},
  volume = {255},
  issue = {-},
  pages = {128181},
  numpages = {14},
  year = {2024},
  month = {-},
  publisher = {Elsevier},
  doi = {10.1016/j.ijbiomac.2023.128181},
}

@article{DAGNINOLEONE20221506,
title = {Phycobiliproteins: Structural aspects, functional characteristics, and biotechnological perspectives},
journal = {Computational and Structural Biotechnology Journal},
volume = {20},
pages = {1506-1527},
year = {2022},
doi = {10.1016/j.csbj.2022.02.016},
author = {Jorge Dagnino-Leone and Cristina Pinto Figueroa and Mónica Latorre Castañeda and Andrea Donoso Youlton and Alejandro Vallejos-Almirall and Andrés Agurto-Muñoz and Jessy {Pavón Pérez} and Cristian Agurto-Muñoz},
}

@article{NewRPE_1996,
title = {The Discovery of a Novel R-phycoerythrin from an Antarctic Red Alga},
journal = {Journal of Biological Chemistry},
volume = {271},
number = {29},
pages = {17157-17160},
year = {1996},
url = {https://www.jbc.org/article/S0021-9258(18)31311-5/fulltext},
author = {Robert MacColl and Leslie E. Eisele and Edwin C. Williams and Samuel S. Bowser},
}

@article{Efrima_1984,
title = {Raman optical activity of molecules adsorbed on metal surfaces: Theory},
journal = {J. Chem. Phys.},
volume = {83},
number = {3},
pages = {1356-1362},
year = {1984},
doi = {10.1063/1.449452},
author = {Efrima S.},
publisher = {AIP},
}

@article{Efrima_1983,
title = {The effect of large electric field gradients on the Raman optical activity of molecules adsorbed on metal surfaces},
journal = {Chemical Physics Letters},
volume = {102},
number = {1},
pages = {79-82},
year = {1983},
doi = {10.1016/0009-2614(83)80662-9},
author = {Efrima S.},
publisher = {Elsevier},
}

@article{Fang_2025,
author = {Gao, Nan and Liu, Haoran and Fang, Yurui},
title = {Coupling Dichroism in Strong-Coupled Chiral Molecule-Plasmon Nanoparticle System},
journal = {The Journal of Physical Chemistry C},
volume = {129},
number = {11},
pages = {5543-5555},
year = {2025},
doi = {10.1021/acs.jpcc.5c00721}
}

@article{MunRho_2019,
title = {Importance of higher-order multipole transitions on chiral nearfield interactions},
author = {Jungho Mun and Junsuk Rho},
pages = {941--948},
volume = {8},
number = {5},
journal = {Nanophotonics},
doi = {doi:10.1515/nanoph-2019-0046},
year = {2019},
}

@article{Kadodwala_2020,
title = {Superchiral near fields detect virus structure},
author = {Tarun Kakkar and Chantal Keijzer and Marion Rodier and Tatyana Bukharova and Michael Taliansky and Andrew J. Love and Joel J. Milner and Affar S. Karimullah and Laurence D. Barron and Nikolaj Gadegaard and Adrian J. Lapthorn and Malcolm Kadodwala},
pages = {195},
volume = {9},
number = {-},
journal = {Light Sci Appl},
doi = {10.1038/s41377-020-00433-1},
year = {2020},
}

@article{AuNP_Peng,
title = {A Facile Synthesis of Monodisperse Au Nanoparticles and Their Catalysis of CO Oxidation},
author = {Sheng Peng and Youngmin Lee and Chao Wang and Hongfeng Yin and Sheng Dai and Shouheng Sun},
pages = {229-234},
volume = {1},
number = {-},
journal = {Nano Research},
doi = {10.1007/s12274-008-8026-3},
year = {2008},
}

@article{Tibaldi2019,
title = {Electrolyte-gated organic field-effect transistors (EGOFETs) as complementary tools to electrochemistry for the study of surface processes},
author = {Alexandra Tibaldi and Laure Fillaud and Guillaume Anquetin and Marion Woytasik and Samia Zrig and Beno\^{i}t Piro and Giorgio Mattana and Vincent No\"{e}l},
pages = {43-46},
volume = {98},
number = {-},
journal = {Electrochemistry Communications},
doi = {10.1016/j.elecom.2018.10.022},
year = {2019},
}

@article{LIU2009,
title = {Probing the pH sensitivity of R-phycoerythrin: Investigations of active conformational and functional variation},
journal = {Biochimica et Biophysica Acta (BBA) - Bioenergetics},
volume = {1787},
number = {7},
pages = {939-946},
year = {2009},
doi = {10.1016/j.bbabio.2009.02.018},
author = {Lu-Ning Liu and Hai-Nan Su and Shi-Gan Yan and Si-Mi Shao and Bin-Bin Xie and Xiu-Lan Chen and Xi-Ying Zhang and Bai-Cheng Zhou and Yu-Zhong Zhang},
}

@article{Kovalev2013,
author = {Bekasova, Olga and Shubin, Vladimir and Safenkova, Irina and Kovalev, Leonid},
year = {2013},
month = {10},
pages = {623-628},
title = {Structural changes in R-phycoerythrin upon CdS quantum dot synthesis in tunnel cavities of protein molecules},
volume = {62},
journal = {International journal of biological macromolecules},
doi = {10.1016/j.ijbiomac.2013.10.010}
}

@article{Rakic95,
author = {Aleksandar D. Raki\'{c}},
journal = {Appl. Opt.},
number = {22},
pages = {4755--4767},
publisher = {Optica Publishing Group},
title = {Algorithm for the determination of intrinsic optical constants of metal films: application to aluminum},
volume = {34},
month = {Aug},
year = {1995},
doi = {10.1364/AO.34.004755},
}

@misc{femcode,
author = {Guillaume Dem{\'e}sy},
title = {Finite element model for crossed grating},
howpublished = {\url{https://gitlab.onelab.info/doc/models/-/tree/master/DiffractionGratings}},
note = {Accessed: 2026-03-06}
}

@article{Demesy10,
author = {Guillaume Dem\'{e}sy and Fr\'{e}d\'{e}ric Zolla and Andr\'{e} Nicolet and Mireille Commandr\'{e}},
journal = {J. Opt. Soc. Am. A},
number = {4},
pages = {878--889},
publisher = {Optica Publishing Group},
title = {All-purpose finite element formulation for arbitrarily shaped crossed-gratings embedded in a multilayered stack},
volume = {27},
month = {Apr},
year = {2010},
url = {https://opg.optica.org/josaa/abstract.cfm?URI=josaa-27-4-878},
doi = {10.1364/JOSAA.27.000878}
}

@article{getdp,
  title={A general environment for the treatment of discrete problems and its application to the finite element method},
  author={Dular, P. and Geuzaine, C. and  Henrotte, F. and Legros, W.},
  journal={IEEE Trans. Mag.},
  volume={34},
  number={5},
  pages={3395--3398},
  year={1998},
  publisher={IEEE}
}

@article{gmsh,
    author = {C. Geuzaine and J.-F. Remacle},
    title  = "Gmsh: a three-dimensional finite element mesh generator with built-in pre- and post-processing facilities",
    journal= "International Journal for Numerical Methods in Engineering",
    volume = "79",
    number = "11",
    pages  = "1309--1331",
    year   = "2009"}

@article{CHANG1996,
title = {Crystal Structure of R-phycoerythrin fromPolysiphonia urceolataat 2.8 A Resolution},
journal = {Journal of Molecular Biology},
volume = {262},
number = {5},
pages = {721-722},
year = {1996},
doi = {10.1006/jmbi.1996.0547},
author = {Wen-rui Chang and Tao Jiang and Zhu-li Wan and Ji-ping Zhang and Zi-xuan Yang and Dong-cai Liang}
}

@Article{Miles2016,
author ="Miles, A. J. and Wallace, B. A.",
title  ="Circular dichroism spectroscopy of membrane proteins",
journal  ="Chem. Soc. Rev.",
year  ="2016",
volume  ="45",
issue  ="18",
pages  ="4859-4872",
publisher  ="The Royal Society of Chemistry",
doi  ="10.1039/C5CS00084J",
}

@article{LinkDNA2019,
author = {Lan, Xiang and Zhou, Xu and McCarthy, Lauren A. and Govorov, Alexander O. and Liu, Yan and Link, Stephan},
title = {DNA-Enabled Chiral Gold Nanoparticle–Chromophore Hybrid Structure with Resonant Plasmon–Exciton Coupling Gives Unusual and Strong Circular Dichroism},
journal = {Journal of the American Chemical Society},
volume = {141},
number = {49},
pages = {19336-19341},
year = {2019},
doi = {10.1021/jacs.9b08797},
URL = {https://doi.org/10.1021/jacs.9b08797},
}

@article{Dionne2013,
author = {Garcia-Etxarri, Aitzol and Dionne, Jennifer A.},
title = {Surface-enhanced circular dichroism spectroscopy mediated by nonchiral nanoantennas},
journal = {Physical Review B},
volume = {87},
number = {},
pages = {235409},
year = {2013},
doi = {10.1103/PhysRevB.87.235409},
}

\newpage

\section{Graphical TOC}
\begin{figure}[H]
\centering\includegraphics[width=8cm]{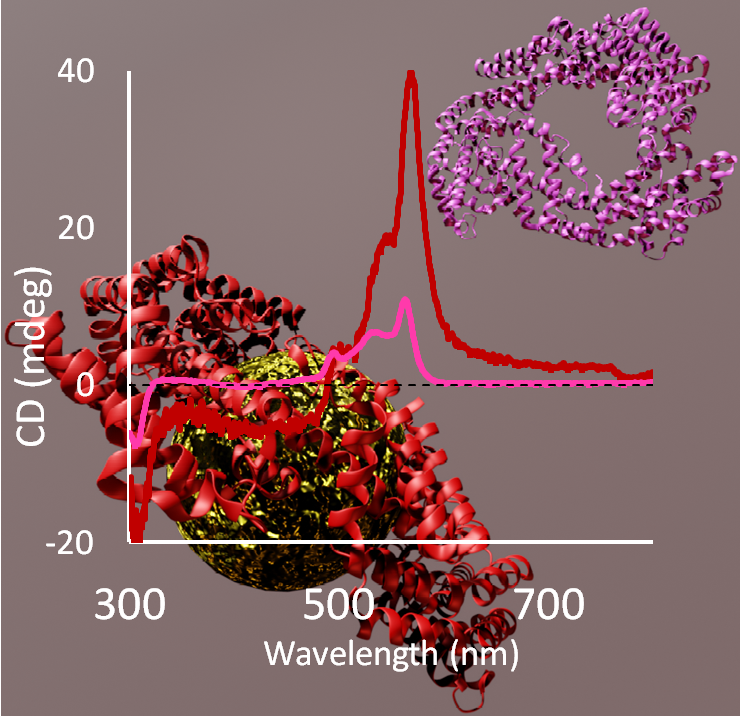}
\end{figure}

\end{document}